\RequirePackage{lineno}
\documentclass[prl,twocolumn,amsmath,amssymb,showpacs,superscriptaddress,nofootinbib,table]{revtex4-1}
\usepackage[table]{xcolor}
\usepackage[dvipdfm,CJKbookmarks=true,unicode,colorlinks,linkcolor=blue,anchorcolor=blue,citecolor=blue,pdfborder={0 0 0}]{hyperref} 
\usepackage{graphicx,epsfig,fancyhdr,amsfonts,amssymb,subfigure,psfrag}
\usepackage{dcolumn} 
\usepackage{bm}
\makeatletter
    \def\CT@@do@color{%
      \global\let\CT@do@color\relax
            \@tempdima\wd\z@
            \advance\@tempdima\@tempdimb
            \advance\@tempdima\@tempdimc
    \advance\@tempdimb\tabcolsep
    \advance\@tempdimc\tabcolsep
    \advance\@tempdima2\tabcolsep
            \kern-\@tempdimb
            \leaders\vrule
                    \hskip\@tempdima\@plus  1fill
            \kern-\@tempdimc
            \hskip-\wd\z@ \@plus -1fill }
\makeatother

\usepackage{multirow}

\newcommand{\BR}{{\cal B}}

\newcommand{\etap}{\eta^\prime}
\newcommand{\rhop}{\rho^\prime}
\newcommand{\jpsi}{J/\psi}

\newcommand{\pip}{\pi^+}

\newcommand{\pin}{\pi^-}

\newcommand{\g}{\gamma}
\newcommand{\ar}{\rightarrow}

\begin{document}
\preprint{}
\sloppy
\title{\boldmath Precision Study of $\etap\ar\g\pip\pin$ Decay Dynamics}

\author{\small
M.~Ablikim$^{1}$, M.~N.~Achasov$^{9,e}$, S. ~Ahmed$^{14}$, X.~C.~Ai$^{1}$, O.~Albayrak$^{5}$, M.~Albrecht$^{4}$, D.~J.~Ambrose$^{44}$, A.~Amoroso$^{49A,49C}$, F.~F.~An$^{1}$, Q.~An$^{46,a}$
% ...
\\
\vspace{0.1cm}
(BESIII Collaboration)\\
\vspace{0.05cm}{\it
$^{1}$ Institute of High Energy Physics, Beijing 100049, People's Republic of China\\
% ...
$^{4}$ Bochum Ruhr-University, D-44780 Bochum, Germany\\
$^{5}$ Carnegie Mellon University, Pittsburgh, Pennsylvania 15213, USA\\
% ...
$^{9}$ G.I. Budker Institute of Nuclear Physics SB RAS (BINP), Novosibirsk 630090, Russia\\
% ...
$^{14}$ Helmholtz Institute Mainz, Johann-Joachim-Becher-Weg 45, D-55099 Mainz, Germany\\
% ...
$^{44}$ University of Rochester, Rochester, New York 14627, USA\\
% ...
$^{46}$ University of Science and Technology of China, Hefei 230026, People's Republic of China\\
% ...
$^{49}$ (A)University of Turin, I-10125, Turin, Italy; (B)University of Eastern Piedmont, I-15121, Alessandria, Italy; (C)INFN, I-10125, Turin, Italy\\
% ...
$^{a}$ Also at State Key Laboratory of Particle Detection and Electronics, Beijing 100049, Hefei 230026, People's Republic of China\\
% ...
$^{e}$ Also at the Novosibirsk State University, Novosibirsk, 630090, Russia\\
% ...
}\vspace{0.4cm}}

\begin{abstract}
  Using a low background data sample of $9.7\times10^{5}$
  $\jpsi\ar\gamma\etap$, $\etap\ar\g\pip\pin$ events, which are 2 orders of magnitude larger than those from the previous experiments, recorded with the
  BESIII detector at BEPCII, the decay dynamics of
  $\etap\ar\g\pip\pin$ are studied with both model-dependent and model-independent approaches. The contributions of $\omega$ and the
  $\rho(770)-\omega$ interference are observed for the first time in the decays $\etap\ar\g\pip\pin$ in
  both approaches. Additionally, a contribution from the box anomaly
  or the $\rho(1450)$ resonance is required in the model-dependent
  approach, while the process specific part of the decay amplitude is
  determined in the model-independent approach.
\end{abstract}

\maketitle

The radiative decay $\etap \to \gamma\pi^+ \pi^-$ is the second most
probable decay mode of the $\etap$ meson with a branching fraction of
$(28.9\pm0.5)$\%~\cite{PDG2016} and is frequently used for tagging
$\etap$ candidates. In the vector meson dominance (VMD)
model~\cite{VMD}, this process is dominated by the decay $\etap \to
\gamma \rho(770)$ (hereafter referred to as $\rho^0$).  In the past,
the dipion mass distribution was studied by several experiments, {\it
  e.g.}, JADE~\cite{jade}, CELLO~\cite{cello}, PLUTO~\cite{pluto},
TASSO~\cite{tasso}, TPC/$\gamma\gamma$~\cite{tpc_gg}, and
ARGUS~\cite{argus}, and a peak shift of about $+20~{\rm MeV}$/$c^2$
for the $\rho^0$ meson with respect to the expected position was
observed. Dedicated studies, using about 2000 $\etap \to \gamma\pi^+
\pi^-$ events, concluded that a lone $\rho^0$ contribution in the
dipion mass spectrum did not describe the experimental
data~\cite{LF_Coll}. This discrepancy could be attributed to a higher
term of the Wess-Zumino-Witten anomaly, known as the box anomaly, in
the chiral perturbation theory (ChPT) Lagrangian~\cite{wzw}. To
determine the ratio of these two contributions, it was suggested to
fit the dipion invariant mass spectrum by including an extra
nonresonant term in the decay amplitude to account for the box
anomaly contribution~\cite{Benayoun:1992ty}. Using a sample of 7490$\pm$180 $\etap$ events,
evidence for the box anomaly
contribution with a 4$\sigma$ significance was reported by
the Crystal Barrel experiment~\cite{CBar}, whereas the observation was
not confirmed 
by the L3 experiment~\cite{Acciarri} using 2123$\pm$53 events.

A recently proposed model-independent
approach~\cite{independent_paper}, based on ChPT and dispersion
theory, relates the $\eta/\etap \to \gamma\pi^+ \pi^-$ decay
amplitudes directly to the $e^+e^-\to \pi^+\pi^-$ process,
which dominates the hadron production cross section at low energies and
gives the largest hadronic contribution to the muon anomalous magnetic
moment~\cite{Jegerlehner:2009ry}.
The amplitudes for $\eta/\etap \to \gamma\pi^+ \pi^-$ therein are given
as a product of the pion vector form factor $F_{V}(s)$ and
a reaction specific part $P(s)$, where $s$ is the $\pip\pin$ invariant
mass squared. The $F_{V}(s)$ term 
is extracted from the $e^+e^-\rightarrow
\pi^+\pi^-$ cross section or from $P$-wave isovector $\pi\pi$ phase shifts. The $P(s)$ term, which can be expanded
into a Taylor series around $s=0$, is expected to be similar for
$\eta$ and $\etap$ decays~\cite{Hanhart:2013vba}, and has been determined
in $\eta$ decays by WASA-at-COSY~\cite{Adlarson:2011xb} and
KLOE~\cite{Babusci:2012ft}, but not yet for $\etap$ decays due to
the limited statistics.

In this Letter, we present a precision measurement of the dipion mass
distribution for the $\etap\to\gamma\pi^+ \pi^-$ process originating
from the radiative decays $J/\psi\to\gamma\etap$ based on
$(1310.6\pm7.0)\times10^{6}$ $\jpsi$ events~\cite{jpsi_events}, which is produced in $e^+e^-$ annihilation,
collected with the BESIII detector~\cite{Ablikim:2009vd}. Both
model-dependent and model-independent approaches are used to
investigate the decay dynamics.

Candidates of $\jpsi\ar\g\etap$, $\etap\ar\g\pip\pin$ are required to
have two charged tracks with opposite charge and at least two photons.
The selection criteria for charged tracks and photon candidates are the same as those in Ref.~\cite{etapto3pi_BESIII}, except for the minimum energy requirement of the photon candidates on the barrel showers, which is 40~${\rm MeV}$ instead of 25~${\rm MeV}$ in this analysis.

\begin{figure}[htbp]
\begin{center}
\includegraphics[width=7.5cm, height=5cm]{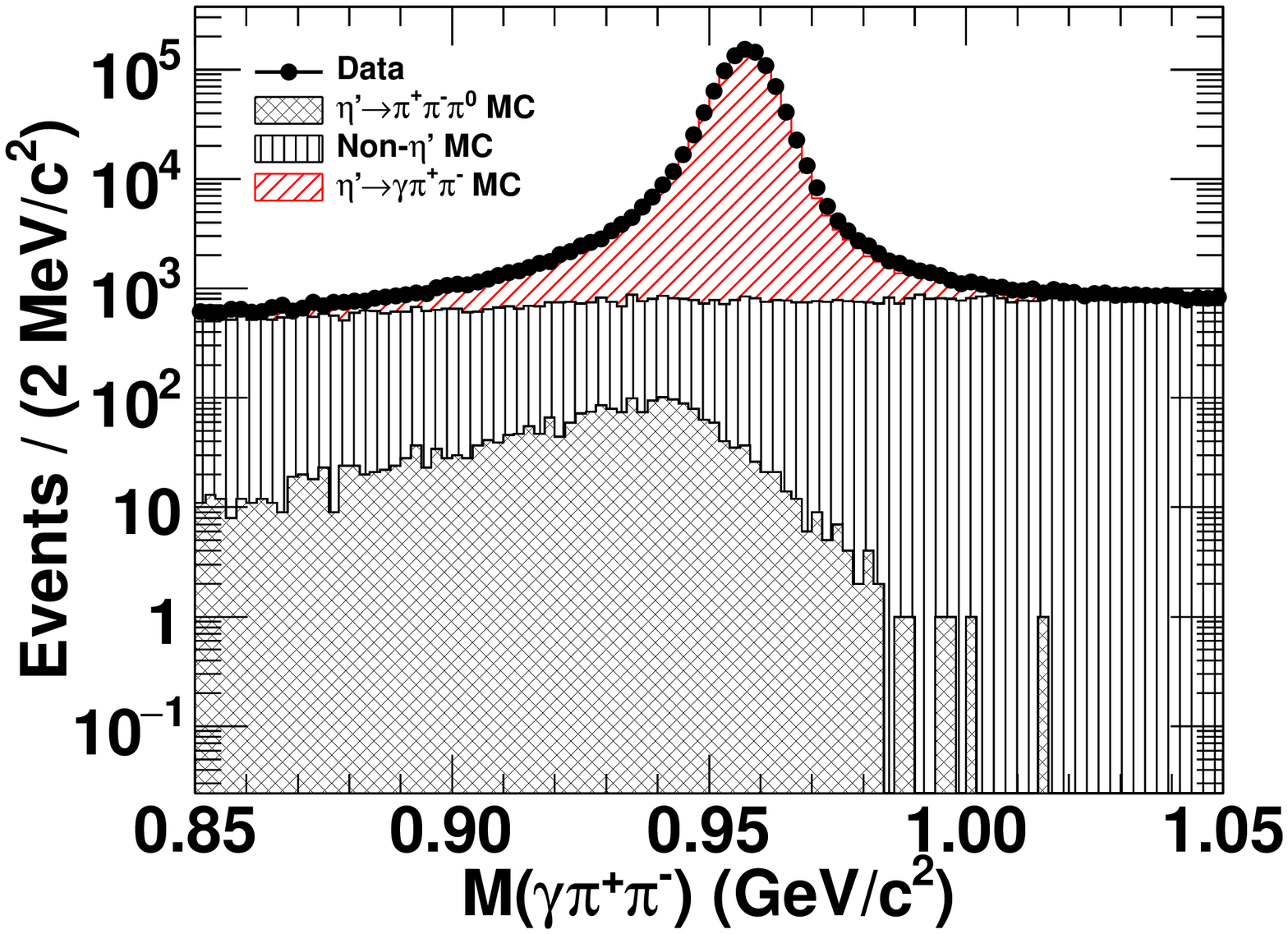}
\caption{Invariant mass spectrum of ${\g\pip\pin}$. Dots with error bars represent the data, and the hatched histograms are MC simulations, where the backgrounds are normalized to the expected contributions as described in the text.}
\label{fig-etapdatamc}
\end{center}
\end{figure}

A four-constraint ($\rm 4C$) energy-momentum conservation kinematic
fit is performed under the $\g\g\pip\pin$ hypothesis, and a loose requirement of $\chi^2_{4C} < 100$ is imposed. This requirement removes 39.3\% background while the efficiency loss is 2.1\%.
For events with
more than two photon candidates, the combination with the smallest
$\chi^2_{4C}$ is retained. In order to remove background events with a
$\pi^0$ in the final states ({\it e.g.},
$J/\psi\rightarrow\pi^+\pi^-\pi^0, \gamma\pi^+\pi^-\pi^0$), we require
that the $\gamma\gamma$ invariant mass is outside the $\pi^0$ mass
region, $|M({\g\g}) - m_{\pi^0}|>0.02$ GeV/$c^2$, where $m_{\pi^0}$ is
the nominal mass of the $\pi^0$~\cite{PDG2016}. Since the radiative photon from the $\eta^\prime$ is always more soft than that from the $J/\psi$ decays, the $\gamma\pi^+\pi^-$
combinations closest to the nominal $\eta^\prime$ mass
($m_{\eta^\prime}$), are kept as $\eta^\prime$ candidates. After the
above selection, a clear $\eta^\prime$ signal is observed in the
$\gamma\pi^+\pi^-$ invariant mass spectrum, as shown in
Fig.~\ref{fig-etapdatamc}. To select candidate events from
$\eta^\prime$ decays, $| M({\gamma\pi^+\pi^-}) -
m_{\eta^\prime}|<0.02$ GeV$/c^2$ is required.

An inclusive Monte Carlo (MC) sample of $1.2\times10^{9}$ $\jpsi$ decay events that are generated with the \textsc{lundcharm} and \textsc{evtgen} models~\cite{Chen:2000, evtgen1} is used to investigate possible background processes. These include events with no $\etap$'s in the final state (non-$\etap$) and those from
$\etap\ar\pip\pin\pi^0$. We use the events in the $\etap$ mass
sideband regions ($0.04 <|M({\g\pip\pin}) - m_{\etap}|<$ 0.06
GeV$/c^2$) to estimate the non-$\etap$ background contribution, which
is at a level of 1.42\%. For the $\etap\ar\pip\pin\pi^0 (\g\g)$
background, a MC study predicts the number of background events to be
0.16\%, 
and its effect is not included in the fit, but taken into
consideration in the systematic uncertainty study.

\begin{figure}[htbp]
\begin{center}
\includegraphics[width=7.5cm, height=5cm]{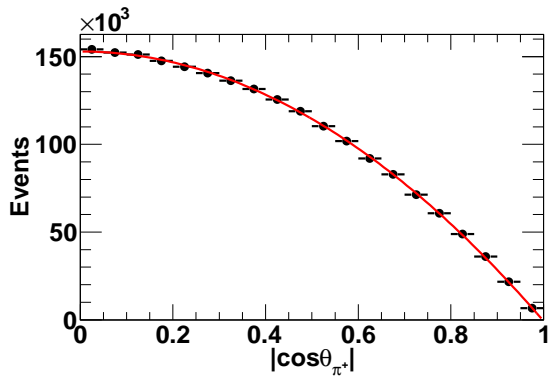}
\caption{Background subtracted and efficiency corrected angular distribution of $\pip$ in the helicity frame of the $\pip\pin$ system. Dots with error bars are data, and the curve is the fit with a $\sin^2\theta_{\pi^+}$ function.}
\label{fig-angular}
\end{center}
\end{figure}

\begin{figure}[htbp]
\centering
\includegraphics[width=7.5cm, height=6.5cm]{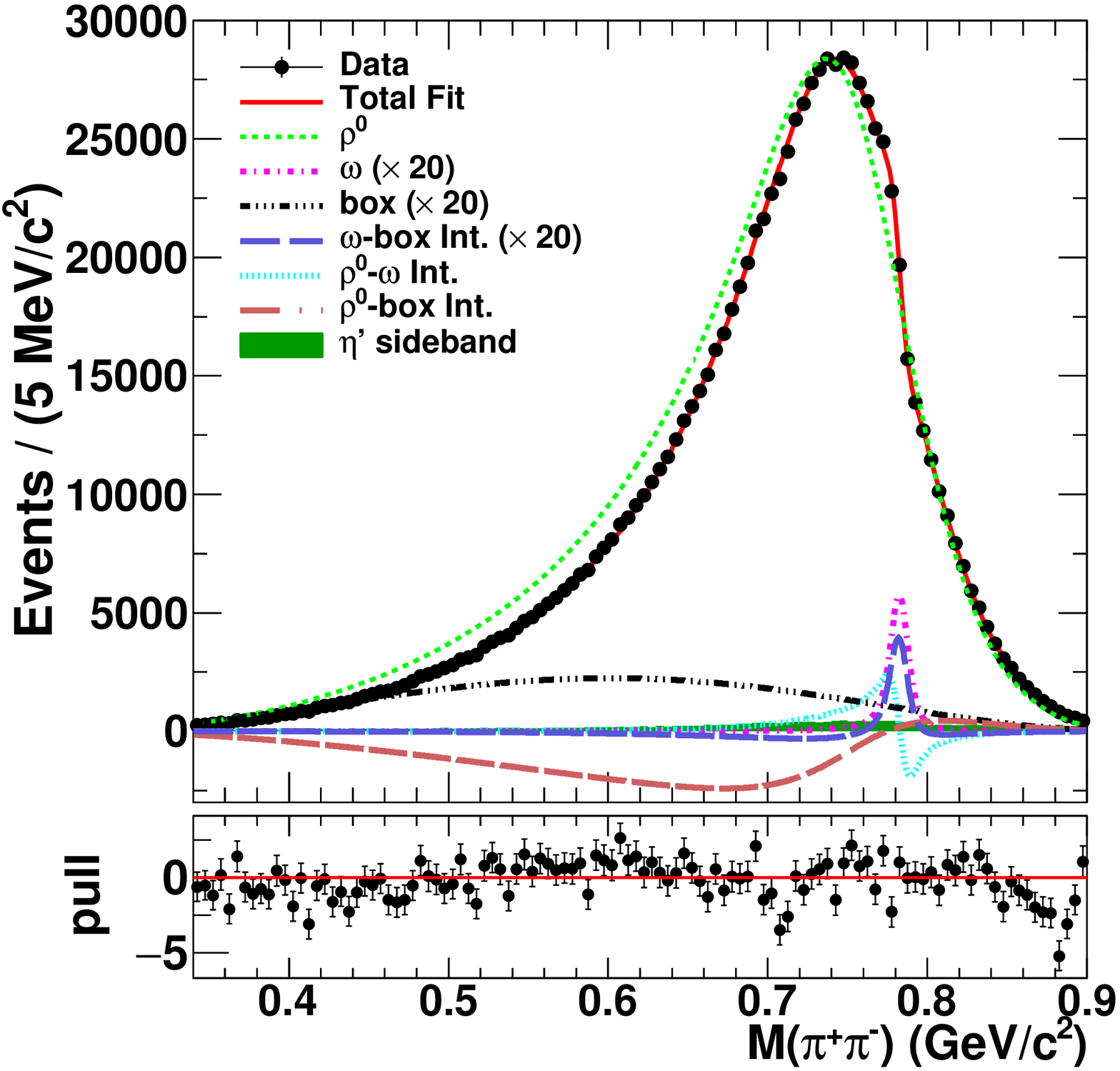}
\put(-40,160){\bf \large~(a)}

\includegraphics[width=7.5cm, height=6.5cm]{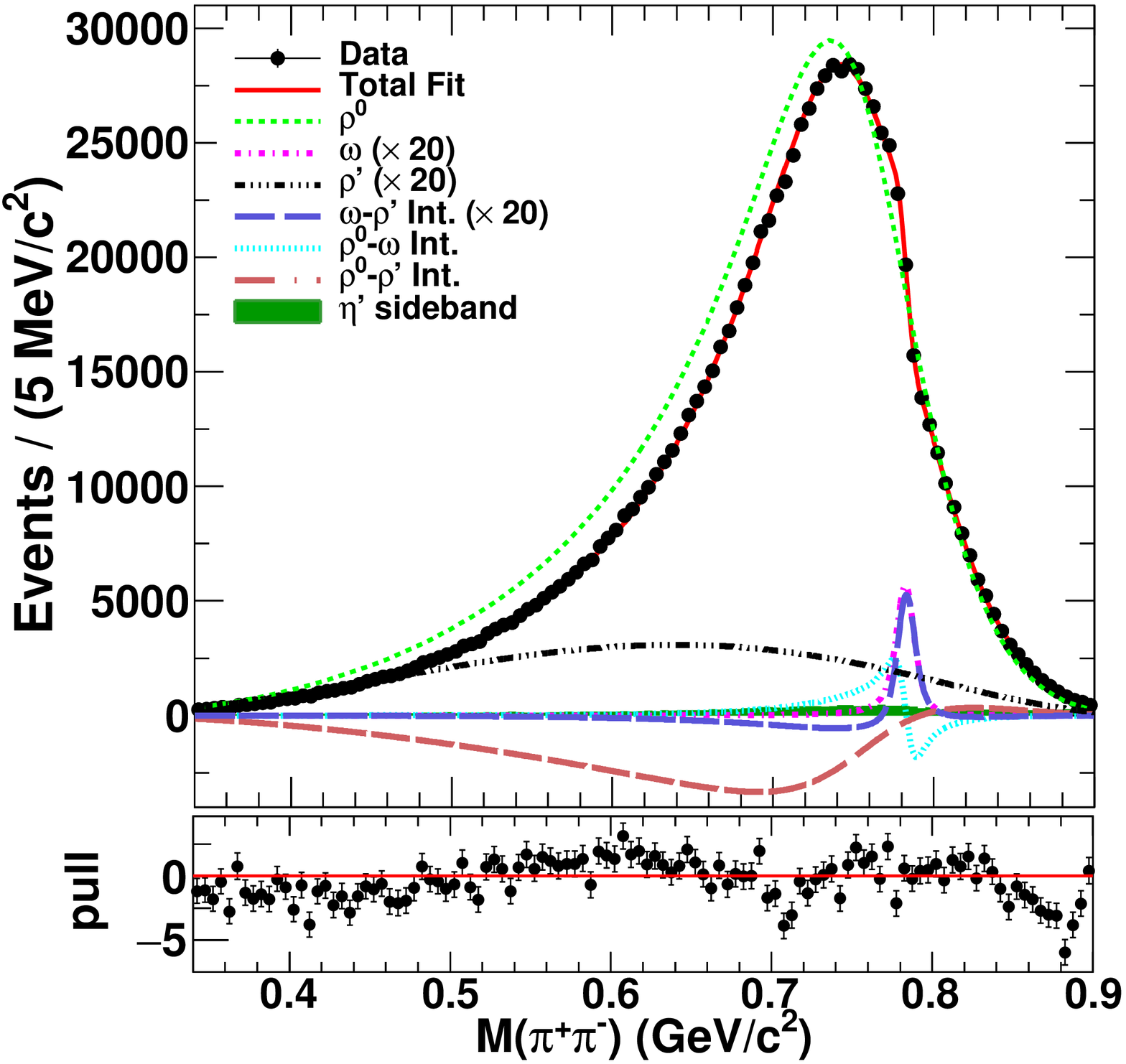}
\put(-40,160){\bf \large~(b)}
\caption{Model-dependent fit results in case (a)
  $\rho^0$-$\omega$-box anomaly and (b)
  $\rho^0$-$\omega$-$\rho^{\prime}$. Dots with error bars represent
  data, the green shaded histograms are the background from $\etap$
  sideband events, the red solid curves are the total fit results,
  and others represent the separate contributions as indicated. To be
  visible, the small contributions of $\omega$, the box anomaly
  ($\rho^{\prime}$) and the interference between $\omega$ and the box
  anomaly ($\rho^{\prime}$) are scaled by a factor of 20.}
\label{fig-fitmass_dependent}
\end{figure}

With the $\etap$ mass window requirement, a low background sample of about $9.7\times10^{5}$
$\etap$ candidates is obtained, which is about 120 times larger than the previous largest sample reported by the Crystal Barrel experiment~\cite{CBar}. 
The background subtracted and
efficiency corrected angular distribution of $\pi^+$ in the helicity
frame of the $\pip\pin$ system, $|\cos\theta_{\pi^+}|$, is shown in
Fig.~\ref{fig-angular}.
The distribution is very well described by
$dN/d\cos\theta_{\pi^+}\propto \sin^2\theta_{\pi^+}$, which is
expected for a $P$-wave dipion system. A detailed MC study indicates
that the reconstructed $\pi^+\pi^-$ invariant mass $M(\pip\pin)$ has a
small shift with respect to the true value, and this is corrected as a
function of $M(\pip\pin)$ according to the values obtained
in MC studies. The maximum shift is less than 0.75 MeV/$c^2$. The
$M(\pip\pin)$ distribution with the mass shift correction is
illustrated as dots with error bars in
Fig.~\ref{fig-fitmass_dependent}.

The dipion mass dependent differential rate is given by~\cite{CBar}
$\frac{d\Gamma}{dM({\pip\pin})}=\frac{k^3_{\gamma}q^3_{\pi}(s)}{48\pi^{3}}|{\cal
  A}|^2$, where $k_{\gamma}=(m_{\etap}^2-s)/(2 m_{\etap})$,
$q_{\pi}(s) = \sqrt{s-4 m_{\pi}^2}/2$ and ${\cal A}$ is the decay
amplitude. Both the model-dependent and model-independent approaches are
carried out to investigate the decay dynamics.

In the model-dependent study, by assuming that the possible
non-$\rho^0$ contributions are from $\omega$, $\rho(1450)$ (hereafter
referred to as $\rho^{\prime}$), and the box anomaly, we
have~\cite{Benayoun:1992ty, CBar, CMD2_ff}
\begin{eqnarray*}
{\cal A}&=  &\frac{BW_{\rho}^{GS}(s)(1+\delta
\frac{s}{M_{\omega}^2}BW_{\omega}(s))+\beta
BW_{\rho^{\prime}}^{GS}(s)}{1+\beta}\nonumber\\
&&\times2\sqrt{48\pi M_{\rho}^{-4}} +\alpha\nonumber,
\label{eq:all}
\end{eqnarray*}
where $\delta$ and $\beta$ are complex numbers representing the
contributions of the $\omega$ and $\rho^{\prime}$ mesons relative to
the $\rho^0$; $\alpha$ is a constant accounting for the box anomaly
contribution~\cite{Benayoun:1992ty}; and $BW_{\rho}^{GS}(s)$,
$BW_{\omega}(s)$, and $BW_{\rho^{\prime}}^{GS}(s)$ are the propagators for
the $\rho^0$, $\omega$, and $\rho^{\prime}$ mesons, respectively. Since
the $\rho^0$ component is dominant in the $M(\pi^+\pi^-)$
distribution, its shape parametrization plays a vital role in the
determination of other components, and is represented with the
Gounaris-Sakurai approach (GS)~\cite{GS_parameter, cmd2}.
$BW_{\omega}(s) = {M_{\omega}^2}/{(M_{\omega}^2 - s - i M_{\omega}
  \Gamma_{\omega})}$, where $M_{\omega}$ and $\Gamma_{\omega}$ are the
$\omega$-meson mass and width, respectively. The $\rho^{\prime}$ is
also described with the GS parametrization. The masses and widths for
the $\omega$ and $\rho^{\prime}$ mesons are fixed to their nominal
values~\cite{PDG2016}, while those for $\rho^0$ are floated in the fit.

Binned maximum likelihood fits are performed to the $M({\pip\pin})$
distribution between 0.34 and 0.90 GeV$/c^2$ with different scenarios,
where the decay amplitude is corrected by a $M(\pip\pin)$-dependent
detection efficiency and is smeared with a $M(\pip\pin)$-dependent
Gaussian function to account for the experimental mass resolution.
The non-$\etap$ background is represented by the $\etap$ sideband
events as discussed above, and is fixed in the fit.  Fits with only
the $\rho^0$ contribution and with additional $\rho^0$-$\omega$
interference give the goodness of fit $\chi^2/ndf$=3365/110 and
3094/108, respectively, where $ndf$ is the number of degrees of freedom.
The results indicate that these components are insufficient
to describe the data and extra contributions are necessary. To improve
the description of the data, we performed a fit, shown in
Fig.~\ref{fig-fitmass_dependent}(a), including the additional box
anomaly term together with $\rho^0$-$\omega$ interference, and much
better agreement with $\chi^2/ndf$=207/107 is obtained. An
alternative fit by replacing the box anomaly with the $\rho^{\prime}$
component gives considerably worse agreement with
$\chi^2/ndf$=303/106, as illustrated in
Fig.~\ref{fig-fitmass_dependent}(b). Fit results of the above two
cases are summarized in Table~\ref{tab-fit_results_syst}. Both cases
yield $\rho^0$ mass and width close to those in the
PDG~\cite{PDG2016}. A fit including both the $\rho^{\prime}$ and box
anomaly gives a reasonable goodness of fit ($\chi^2/ndf$
=134/105). 
However, a very strong correlation in amplitude between the
box anomaly and the $\rhop$ components, $i.e.$ the correlation
coefficient is -0.986, is observed, due to the tail of the
$\rho^{\prime}$ having a similar line shape as that of the box anomaly. Thus they are not well under control, and it is hard for one to distinguish them in the fitting. Whereas the mass and width of the $\rho^0$ are stable, which are $776.43\pm0.36$, $150.26\pm0.56$ MeV/$c^2$, respectively.
Therefore a refined model dependent amplitude beyond
including just the $\rhop$ or the box anomaly contribution is desirable.

\begin{table*}[htbp]
   \centering
\caption{The results of the model-dependent fits to the $M({\pip\pin})$ distribution in different cases. The first uncertainties are statistical and the second ones systematic.}
\label{tab-fit_results_syst}
        \begin{tabular}{lp{4.5cm}p{4.5cm}}
        \hline
Model-dependent fit                          &~~~$\rho^0$-$\omega$-box anomaly     &~~~~~~~~~~~$\rho^0$-$\omega$-$\rho^{\prime}$          \\
       \hline
       $M(\rho^{0})$ [MeV/c$^2$]       &774.34~$\pm$~0.18~$\pm$~0.35         &~772.93~$\pm$~0.18~$\pm$~0.34               \\
       $\Gamma(\rho^{0})$ [MeV]       &150.85~$\pm$~0.55~$\pm$~0.67         &~150.18~$\pm$~0.55~$\pm$~0.65                 \\
       arg $\delta$ [rad]                         &~~(0.65~$\pm$~3.14~$\pm$~2.62)$\times10^{-2}$     &~~(-2.59~$\pm$~3.19~$\pm$~2.62)$\times10^{-2}$           \\
       $|\delta|$ [$10^{-3}$]                 &~~~1.61~$\pm$~0.05~$\pm$~0.13        &~~~~1.59~$\pm$~0.05~$\pm$~0.11                   \\
      arg $\beta$ [rad]                           &~~~...                                                             &~~~~3.28~$\pm$~0.11~$\pm$~0.04              \\
      $|\beta|$                                        &~~~...                                                             &~~~~0.26~$\pm$~0.01~$\pm$~0.01              \\
$\alpha$ [MeV$^{-2}]$                     &-11.56~$\pm$~0.21~$\pm$~0.32            &~~~~...                                                   \\\hline

$\BR(\eta^{\prime}\rightarrow\g\rho^0)$    &(33.34~$\pm$~0.06$~\pm$~1.60)\%  &~~(34.43~$\pm$~0.52$~\pm$~1.97)\%                \\

$\BR(\eta^{\prime}\rightarrow\g\omega\to\g\pip\pin)$ &~(3.25~$\pm$~0.21~$\pm$~0.52)$\times10^{-4}$   &~~~(3.22~$\pm$~0.21$~\pm$~0.52)$\times 10^{-4}$      \\
$\BR(\eta^{\prime}\rightarrow\g\pip\pin$ via box) &~(2.45~$\pm$~0.09~$\pm$~0.19)$\times10^{-3}$          &~~~~...  \\
$\BR(\eta^{\prime}\rightarrow\g\pip\pin$ via $\rho^{\prime})$  &~~~...               &~~~(3.43~$\pm$~0.38~$\pm$~0.28)$\times10^{-3}$   \\  \hline \\
         \end{tabular}
\end{table*}

As suggested by Ref.~\cite{independent_paper}, a model independent
approach is also implemented to investigate the decay dynamics. The
decay amplitude follows ${\cal A}=N P(s) F_{V}(s)$, where
$N$ is a normalization factor, a polynomial function $P(s) = 1 +
\kappa s + \lambda s^2+ \xi 
BW_{\omega}+\mathcal{O}(s^4)$ includes the possible $\omega$ term
$\xi$ and quadratic term $\lambda$, and the pion vector form factor
$F_{V}(s)$ is obtained from $e^+e^-\ar \pip \pin$
measurements~\cite{bes3_crossPipi,ff_papers,ff_babar,ff_kloe,ff_na7}.

\begin{figure}[htbp]
\includegraphics[width=7.5cm, height=6.5cm]{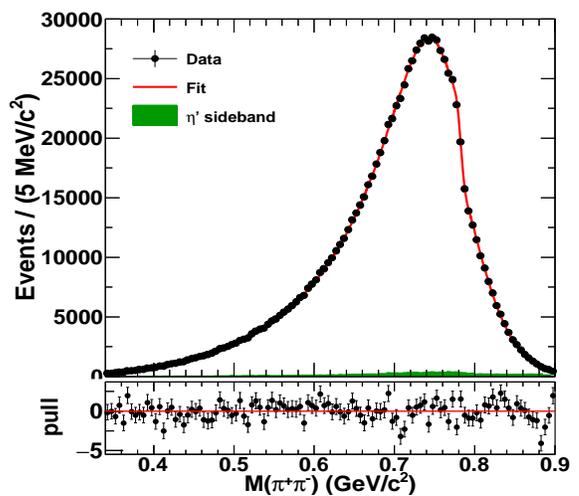}
\caption{The results of the model-independent fit with
  $\omega$ interference. Dots with error bars represent data, the
  (green) shaded histogram is the background contribution from $\etap$
  side-band events, and the (red) solid curve is the fit result.}
\label{fig-independent_fit}
\end{figure}

A fit to the data gives $\kappa=0.992~\pm~0.039$ GeV$^{-2}$,
$\lambda=-0.523~\pm~0.039$ GeV$^{-4}$, $\xi=0.199~\pm~0.006$ with
$\chi^2/ndf$=145/109, where the uncertainties are statistical
only. The fit result is shown in Fig.~\ref{fig-independent_fit}, and
the statistical significances of nonzero quadratic term and $\omega$
term are $13\sigma$ and $34\sigma$, respectively, which are estimated
with the changes of the log likelihood value and the number of degree
of freedoms. An alternative fit without the $\omega$ contribution
yields $\kappa=1.420~\pm~0.047$ GeV$^{-2}$ and
$\lambda=-0.951~\pm~0.046$ GeV$^{-4}$, which is compatible to a recent
prediction $\lambda=-1.0\pm0.1$
GeV$^{-4}$~\cite{Kubis:2015sga}. However, this fit corresponds to a
very poor goodness of fit ($\chi^2/ndf$ =1351/110) and fails to
describe the data. Different from the measurements of
$\eta\ar\g\pip\pin$ decays~\cite{Adlarson:2011xb, Babusci:2012ft},
which are not sensitive to the quadratic term, both the quadratic term
and the $\omega$ contribution are significant in the
$\etap\ar\g\pip\pin$ decays.

The systematic uncertainties in the model-dependent and model-independent
approaches are discussed in detail in the following and are summarized
in the Supplemental Material~\cite{SM_tab-syst_error}.
The total systematic uncertainty is
the quadrature sum of the individual values by assuming them to be
independent.

The uncertainty associated with the 4C kinematic fit originates from
the difference between data and MC simulation. This difference is
reduced by correcting the track helix parameters of the MC sample as
described in Ref.~\cite{track_helix}. To estimate the corresponding
uncertainty, the analysis is repeated without the track helix
parameters correction, and the resultant change is assigned as the
uncertainty.

The MDC tracking and photon detection efficiencies are studied based
on a clean sample of $J/\psi\rightarrow\rho\pi$. The differences
between data and MC simulation are investigated as a function of
momentum (energy), and are less than 1\% for each charged track and
1\% for each photon~\cite{tracks_syst}. To evaluate their impact on
the results, an event-by-event correction on the tracking and photon
detection efficiency is performed as a function of momentum (energy).
The resultant changes on the results are taken as the systematic
uncertainties.

The uncertainty from the $\etap$ mass window requirement is evaluated
by varying the required values by $\pm$ 6 MeV/$c^2$, which is the mass
resolution from the MC simulation, and the maximum change of the
results is taken as the uncertainty.

Systematic sources related with the fit procedure include the binning,
the fit range, the background, the mass resolution of $M(\pip\pin)$,
and the input parameters in the fit. The uncertainty from binning is
studied with the same fit procedure with varied bin width.  For the
uncertainty due to the fit range, we 
take the larger change of
the fit result with varied fit ranges as the uncertainty. Two systematic sources, $i. e.$
the $\etap$ sideband and the small contribution of
$\etap\ar\pip\pin\pi^0$, are considered as the uncertainty
related with the background in the fit. The former one is estimated by
changing the sideband region, while the latter one is studied by
including the background in the fit with a fixed magnitude and shape
in accordance with the MC study. We assign the quadratic sum of the
two uncertainties as the total background uncertainty. The impact
caused by the $\pip\pin$ mass resolution is estimated by varying the
resolution by $\pm$10\% in the fit, and the maximum change of the fit
result is assigned as the uncertainty. For the model dependent study,
the uncertainty due to the mass and width of $\omega$, $\rhop$
resonances is estimated by varying the input values with $\pm1\sigma$
of the corresponding uncertainties from the PDG~\cite{PDG2016},
respectively, and taking the quadratic sum of the maximum change of
the fit results as the uncertainty of the resonance parameters.

For the measurement of the branching fraction of $\eta^\prime$ decays into
$\gamma\rho^0$, $\gamma\omega$, $\g$ box anomaly and
$\gamma\rho^{\prime}$, the additional uncertainties from the branching
fractions of $J/\psi\rightarrow\gamma\eta^\prime$~\cite{PDG2016} and
the number of $\jpsi$ events~\cite{jpsi_events} are also taken into
account.

In the model independent approach, the uncertainty associated with the
input pion vector form factor $F_{V}(s)$, is estimated by an
alternative fit incorporating the line shape of $F_{V}(s)$ from
Ref.~\cite{FF_syst}. The resulting differences, 16.4\%, 34.7\%, and
3.4\% for the $\kappa$, $\lambda$, $\xi$ parameters, respectively,
determine the systematic uncertainty. Since this uncertainty is
theoretically dependent, it is treated as a separated uncertainty in
the final results.

In summary, the $\etap\ar\g\pip\pin$ decay dynamics is studied based
on a sample of $9.7\times10^5$ events originating from the radiative
decay $J/\psi\rightarrow\gamma\eta^\prime$ of
$1.31\times10^9$ $\jpsi$
events collected with the BESIII
detector. We have measured the dipion invariant mass distribution and
performed fits using model dependent and independent approaches.  For
the first time, the $\omega$ contribution is observed in the dipion
mass spectrum in the decays $\etap\ar\g\pip\pin$. The model-dependent
fit indicates that only the components of $\rho^0$ and $\omega$ as
well as the corresponding interference fail to describe the data, and
an extra significant contribution, $i.e.$ the box anomaly or $\rhop$,
is found to be necessary for the first time. The corresponding fit
results and the measured branching fractions are summarized in
Table~\ref{tab-fit_results_syst}. The data call for a more complete
model-dependent amplitude beyond just including the box anomaly or
$\rhop$ contribution for the $M(\pip\pin)$ spectrum.

The model independent approach~\cite{independent_paper} provides a
satisfactory parametrization of the dipion invariant mass spectrum,
and yields the parameters of the process-specific part $P(s)$ to be
$\kappa=0.992\pm0.039\pm0.067\pm0.163$ GeV$^{-2}$,
$\lambda=-0.523\pm0.039\pm0.066\pm0.181$ GeV$^{-4}$, and
$\xi=0.199\pm0.006\pm0.011\pm0.007$, where the first uncertainties are
statistical, the second are systematic, and the third are
theoretical. In contrast to the conclusion in
Ref.~\cite{independent_paper} based on the limited statistics from the
Crystal Barrel experiment~\cite{CBar}, our result indicates that the
quadratic term and the $\omega$ contribution in $P(s)$, corresponding
to statistical significances of $13\sigma$ and $34\sigma$,
respectively, are necessary. 

The BESIII Collaboration thanks the staff of BEPCII and the IHEP computing center for their strong support. This work is supported in part by National Key Basic Research Program of China under Contract No. 2015CB856700; National Natural Science Foundation of China (NSFC) under Contracts No. 11565006, No. 11235011, No. 11335008, No. 11425524, No. 11625523, No. 11635010, No. 11675184, No. 11735014; the Chinese Academy of Sciences (CAS) Large-Scale Scientific Facility Program; the CAS Center for Excellence in Particle Physics (CCEPP); Joint Large-Scale Scientific Facility Funds of the NSFC and CAS under Contracts No. U1332201, No. U1532257, No. U1532258; CAS Key Research Program of Frontier Sciences under Contracts No. QYZDJ-SSW-SLH003, No. QYZDJ-SSW-SLH040; 100 Talents Program of CAS; National 1000 Talents Program of China; Shandong Natural Science Funds for Distinguished Young Scholar under Contract No. JQ201402; INPAC and Shanghai Key Laboratory for Particle Physics and Cosmology; German Research Foundation DFG under Contracts Nos. Collaborative Research Center CRC 1044, FOR 2359; Istituto Nazionale di Fisica Nucleare, Italy; Koninklijke Nederlandse Akademie van Wetenschappen (KNAW) under Contract No. 530-4CDP03; Ministry of Development of Turkey under Contract No. DPT2006K-120470; National Natural Science Foundation of China (NSFC) under Contracts No. 11505034, No. 11575077; National Science and Technology fund; The Swedish Research Council; U. S. Department of Energy under Contracts No. DE-FG02-05ER41374, No. DE-SC-0010118, No. DE-SC-0010504, No. DE-SC-0012069; University of Groningen (RuG) and the Helmholtzzentrum fuer Schwerionenforschung GmbH (GSI), Darmstadt; WCU Program of National Research Foundation of Korea under Contract No. R32-2008-000-10155-0.

\end{document}